\documentclass[twocolumn,aps,pre,amsmath,showpacs]{revtex4}
\usepackage{graphicx}

\def\be{\begin{equation}}
\def\ee{\end{equation}}
\def\bc{\begin{center}}
\def\ec{\end{center}}
\def\lan{\langle}
\def\ran{\rangle}
\def\r{{\vec r}}
\def\k{{\vec k}}

\begin{document}

\title{Dynamical heterogeneity in a model for permanent gels: Different behavior of dynamical susceptibilities}
\author{T. Abete$^{a,b}$, A. de Candia$^{a,b,c}$, E. Del Gado$^{d}$,
A. Fierro$^{e}$, and A. Coniglio$^{a,c,e}$} \affiliation{${}^a$
Dipartimento di Scienze Fisiche, Universit\`a di Napoli ``Federico
II'',\\ Complesso Universitario di Monte Sant'Angelo, via Cintia
80126 Napoli, Italy} \affiliation{${}^b$ CNISM Universit\`a di
Napoli ``Federico II''} \affiliation{${}^c$ INFN Udr di Napoli}
\affiliation{${}^d$ ETH Z\"urich, Department of Materials, Polymer
Physics, CH-8093 Z\"urich, Switzerland} \affiliation{${}^e$ INFM CNR
Coherentia}
\date{\today}

\begin{abstract}
We present a systematic study of dynamical heterogeneity in a model
for permanent gels, upon approaching the gelation threshold. We find
that the fluctuations of the self intermediate scattering function
are increasing functions of time, reaching a plateau whose value, at
large length scales, coincides with the mean cluster size and
diverges at the percolation threshold. Another measure of dynamical
heterogeneities, i.e. the fluctuations of the self-overlap, displays instead
a peak and decays to zero at long times. The peak, however,
also scales as the mean cluster size. Arguments are given for this
difference in the long time behavior. We also find that non-Gaussian
parameter reaches a plateau in the long time limit. The value of the
plateau of the non-Gaussian parameter, which is connected to the
fluctuations of diffusivity of clusters, increases with the volume
fraction and remains finite at percolation threshold.

\end{abstract}


\maketitle
\section{Introduction}
In the context of the glass transition the concept of dynamical
heterogeneities has been very fecund
\cite{cicerone,kob_PRL_97,franz,glotzer,bennemann,berthier,new1,new2,new3,new4,new5,new6}.
In glassy systems the correlated motion of particles manifest as
significant fluctuations around the average dynamics,
strongly increasing
as the transition is approached. These heterogeneities in the
dynamics have been studied quantitatively via the so-called
dynamical susceptibility \cite{franz}, $\chi_4(t)=N(\langle
F^2(t)\rangle-\langle F(t)\rangle^2)$, obtained as the fluctuations
of a suitable time dependent correlator $F(t)$ (where $N$ is the
number of particles and $\langle \dots \rangle$ is the ensemble
average). Two quantities are usually considered: The fluctuations of
the self intermediate scattering functions (ISF)
\cite{biroli,berthier}
$\chi_4(k,t)=N\left[\langle|\Phi_s(k,t)|^2\rangle-\langle\Phi_s(k,t)\rangle^2\right]$
usually measured in numerical simulations, and the fluctuations of
the time dependent overlap \cite{59,60,61} $\chi_4^Q(t)= N [\langle
q(t)^2\rangle-\langle q(t)\rangle^2]$,
which, first introduced in $p$-spin glass models \cite{franz}, has
been calculated also within mode coupling theory
\cite{franz,biroli}. For the fluctuations of the overlap, the role
of the inverse of the wave vector $k$ is played by the parameter $a$
characterizing the overlap function, which is different from zero
only if a particle has moved a distance less than the fixed value
$a$. In usual glassy systems the behavior observed in the dynamical
susceptibility is essentially the same despite of different choices
of $F(t)$ \cite{glotzer,biroli,berthier}: $\chi_4(t)$ grows as a
function of the time, reaches a maximum and then decreases to a
constant, consistently with the transient nature of the dynamical
heterogeneities. Some differences in the $k$ dependence of these two
quantities were however found in a model for glasses
\cite{chandler_garr}.

Recently, dynamical heterogeneities have been studied in other complex
systems, such as granular media
\cite{noi_granulari,berthier_gra,dauchot,glotzer_gra} and attractive colloidal
systems \cite{weeks, cipelletti,attr_sim, reichman, charb}, where
behaviors qualitatively similar to that found in glasses are
observed. In particular, Ref.\cite{charb} reports a systematic study of
the dynamic susceptibility
in colloidal systems along the attractive glassy line.
Typically the dynamical susceptibility, defined as the fluctuations
of the self ISF, displays a well pronounced peak. However, in the
attraction-dominated limit, the dependence on both time and wave
vector markedly differs from that in standard repulsion-dominated
systems (hard-sphere limit).

In a recent letter \cite{tiziana_prl} we have studied the behavior
of the dynamical susceptibility, $\chi_4$, defined as the
fluctuations of the self ISF, in a model for permanent gel, where
bonds are modeled using a finitely extendable non linear elastic
(FENE) potential \cite{FENEdum,FENE} between neighboring particles.
It was found that the behaviour of $\chi_4(k,t)$ is drastically
different from that found in glasses. In fact it grows in time until
it reaches a plateau in the limit of large time $t$, without
decaying to $1$. The value of the plateau in the limit of low wave
vector, $k\rightarrow 0$, was in fact found to coincide with the
mean cluster size. As a consequence, as the system approaches the
gel transition (i.e. the percolation threshold), the value of the
plateau diverges. For a fixed value of $k$, the value of the plateau
coincides with the mean cluster size up to
a linear size of the order of the inverse of $k$. Therefore, for any
$k >0$ ($k>k_{min}$ in our study), the plateau never diverges: it
decreases as $k$ increases and eventually goes to one.

Here we present a systematic study of this
FENE model for permanent gels \cite{tiziana_prl}. Moreover, we
compare the behavior of the fluctuations of the self ISF and of the
self-overlap, and find a marked difference between the two ones. The
first one, as mentioned above, is an increasing function of the time
and tends to a plateau, whereas the second one reaches a maximum and
then decreases. However, the value of the maximum scales as the value
of the plateau of the fluctuations of the self ISF, with the same
critical exponent of the mean cluster size.

The reason why these two quantities differ so drastically in the
long time limit is the following: the fluctuation of the overlap is
related to the correlations between the event that a monomer has
moved a distance less then $a$ in a time interval $t$ and the event
that another monomer has also moved a distance less then $a$ in the
same interval $t$. In the long time $t$ all particles have moved a
distance larger than $a$ therefore such correlations are zero. On
the other hand in the long time limit the fluctuation of the self
ISF is related to the correlation of the distance separating monomers
$i$ and $j$ at time $0$ and the distance between the same monomers
at time $t$. This quantity is different from zero if the particles
$i$ and $j$ are in the same cluster.

Although the long time limit of the two quantities $\chi_4(k,t)$ and
$\chi^Q_4(a,t)$ is different in gels with permanent bonds, they have
in common not only the property that plateau and maximum scale in
the same way, but also one key feature which is the strong length
scale dependence: The peak of the fluctuations of the self part of
the overlap and the plateau of the fluctuations of the self ISF
decreases strongly as the wave vector $k$ (or $1/a$) increases,
which is the sign that clusters of bonded particle dominate the
dynamics. The same feature is also valid for strong colloidal gels
\cite{decandia_prep}. This strong length scale dependence of the
dynamical susceptibility seems to be the distinctive sign of
permanent or strong colloidal gelation, compared with the
(attractive or repulsive) glass transition.

Finally, we measure the non-Gaussian parameter $\alpha_2$ and find
that, due to the presence of clusters, it is different from zero
also in the long time limit. However, its plateau value, which is
connected to diffusivity,
remains finite upon approaching the transition.

In Sect.\ref{model} we introduce the model used and give the details
of the numerical simulations. We analyze the self ISF and its
fluctuations in Sect.\ref{sec:self} and the self-overlap and its
fluctuations in Sect.\ref{sec:overlap}. The mean squared
displacement and the non-Gaussian parameter are discussed in
Sect.\ref{diffusion}, whereas
the Sect.\ref{conclusion} contains the concluding remarks. Finally,
in Appendix \ref{gelation} we investigate the static properties of
the sol-gel transition, corresponding to the percolation of
permanent bonds between particles \cite{flo,deg}.

\section{Model and numerical simulations}
\label{model} We consider a $3d$ system of $N$ particles interacting
with a soft potential given by Weeks-Chandler-Andersen (WCA) potential
\cite{chandler}:
\begin{equation}
U_{ij}^{WCA}=\left\{ \begin{array}{ll}
4\epsilon[(\sigma/r_{ij})^{12}-(\sigma/r_{ij})^6+\frac{1}{4}], & r_{ij}<2^{1/6}\sigma \\
0, & r_{ij}\ge2^{1/6}\sigma \end{array} \right.
\end{equation}
where $r_{ij}$ is the distance between the particles $i$ and $j$.

After the equilibration, particles
distant less than $R_0$ are linked by adding an attractive potential:
\begin{equation}
U_{ij}^{FENE}=\left\{ \begin{array}{ll}
-0.5 k_0 R_0^2 \ln[1-(r_{ij}/R_0)^2], & r_{ij}< R_0\\
\infty, & r_{ij}\ge R_0 \end{array} \right.
\end{equation}
representing a finitely extendable nonlinear elastic (FENE). The
FENE potential was firstly introduced in Ref.\cite{FENEdum} and is
widely used to study linear polymers \cite{FENE}. We choose
$k_0=30\epsilon/\sigma^2$ and $R_0=1.5\sigma$ as in Ref.\cite{FENE}
in order to avoid any bond crossing and to use an integration time
step $\Delta t$ not too small \cite{nota}. The introduction of the
FENE potential leads to the formation of permanent bonds among all
the particles whose distance at that time is smaller than $R_0$.

We have performed molecular dynamics simulations of this model: The
equations of motion were solved in the canonical ensemble (with a
Nos\'e-Hoover thermostat) using the velocity-Verlet algorithm
\cite{Nose-Hoover} with a time step $\Delta t=0.001\delta\tau$,
where $\delta\tau=\sigma(m/\epsilon)^{1/2}$ is the standard unit
time for a Lennard-Jones fluid and $m$ is the mass of particle. We
use reduced units where the unit length is $\sigma$, the unit energy
is $\epsilon$ and the Boltzmann constant $k_B$ is set equal to $1$.
We use periodic boundary conditions, and average all the
investigated quantities over $32$ independent configurations of the
system.

The temperature is fixed at $T=2$ and the volume fraction
$\phi=\pi\sigma^3N/6L^3$ (where $L$ is the linear size of the
simulation box in units of $\sigma$) is varied from $\phi=0.02$ to
$\phi=0.12$.  Using the percolation approach, we identify the gel
phase as the state in which there is a percolating cluster
\cite{flo,deg}. A finite size scaling analysis is presented in the
Appendix, showing that this transition is in the universality class
of random percolation. We find that the threshold is $\phi_c =
0.09\pm0.01$. In particular, we obtain that the cluster size
distribution, $n_s\sim s^{-\tau}$ for $\phi=\phi_c$ with
$\tau=2.1\pm0.2$, the mean cluster size $S(\phi)= \sum s^2 n_s /
\sum s n_s\sim (\phi_c-\phi)^{-\gamma}$ with $\gamma=1.8\pm0.1$, and
the connectedness length $\xi\sim(\phi_c-\phi)^{-\nu}$ with
$\nu=0.88\pm0.01$. In the following we fix the number of particles,
$N=1000$.

Due to the introduction of bonds, spatial correlations appear at low
wave vectors. Although these correlations increase as a function of
the volume fraction, the low $k$ limit of the static structure
factor $S(k)$ is always small compared to the number of particles,
and no phase separation is observed.

\section{Self Intermediate Scattering Function and its fluctuations}
\label{sec:self} Relevant information on the relaxation dynamics
over different length scales can be obtained from the self
Intermediate Scattering Functions (ISF) $F_s(k,t)$:
\begin{equation}
F_s(k,t)=\left[\lan \Phi_s(k,t)\ran\right]
\end{equation}
where $\langle \dots \rangle$ is the thermal average over a fixed
bond configuration, $\left[...\right]$ is the average over $32$
independent bond configurations of the system, and
\begin{equation}
\Phi_s(k,t)=\frac{1}{N}\sum_{i=1}^N e^{i\vec{k}\cdot(\vec{r}_i(t)-
\vec{r}_i(0))}
\end{equation}

\begin{figure}
\begin{center}
\includegraphics[width=7cm]{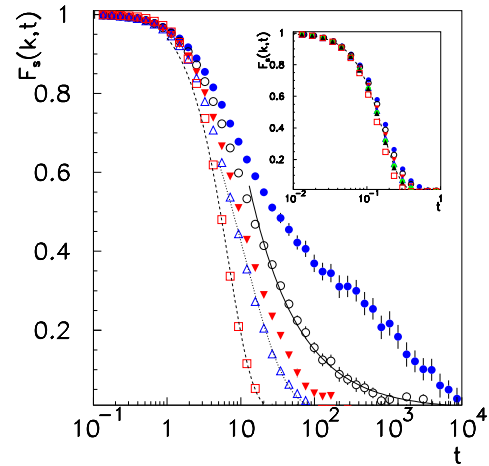}
\caption{(Color online)
{\bf Main frame}: Self ISF for $\phi=0.02$, $0.07$, $0.08$,
$0.09$, $0.1$ (from bottom to top) and $k\sim0.35$ as a function of
time $t$. The lines are fitting curves: For $\phi<<\phi_c$ the decay
is well fitted by an exponential behavior (dashed line); if $\phi$
approaches to $\phi_c$ a stretched exponential decay appears, with
$\beta=0.75\pm0.01$ for $\phi=0.07$ (dotted line). For $\phi=0.09$
the decay is well fitted by a power law $\sim t^{-c}$ with
$c=0.65\pm0.03$ (full line).
 {\bf Inset}: Self ISF for $k\sim7$
and the same volume fractions of main frame.}
\label{fig_self}
\end{center}
\end{figure}

In Fig.\ref{fig_self}, $F_s(k,t)$  is plotted as a function of $t$
for different $\phi$, respectively for $k_{min}=2\pi/L\sim 0.35$
(main frame) and $k\sim 7$ (inset). At the smallest wave vector, for
very low values of the volume fraction, the self ISF decays to zero
following an exponential behavior. As the volume fraction is
increased towards the percolation threshold, we observe the onset of
a stretched exponential decay, $e^{-{(t/\tau)}^\beta}$, with $\beta$
decreasing as a function of $\phi$ (for instance $\beta=0.75\pm0.01$
for $\phi=0.07$ and $\beta=0.58\pm0.02$ for $\phi=0.085$). The
cluster size distribution has  started to widen towards the
percolation regime (see Appendix), and therefore, over sufficiently
large length scales, the behavior of $F_s(k,t)$ is due to the
contribution of different relaxation processes, characterized by
different relaxation times, whose superposition produces a
detectable deviation from an exponential law. Near the transition
threshold the long time decay is characterized by a power law
behavior, indicating that the relaxation over this length scale is
controlled by the formation of the percolating cluster, with a
critically growing relaxation time. If the volume fraction increases
further, the decay becomes slower and slower, showing a logarithmic
behavior. These features of the dynamics well reproduce the
experimental findings \cite{gel_dyn_exp}. Moreover they agree with
results obtained via numerical simulations of different gelation
models \cite{cubetti,ema_EPJ,spike}. At large wave vectors (see
Inset of Fig.\ref{fig_self}) and low volume fractions, $F_s(k,t)$
decays to zero as $e^{-(t/\tau)^{2}}$ (continuous curves in figure),
corresponding to the ballistic regime of particle motion.

\begin{figure}
\begin{center}
\includegraphics[width=7cm]{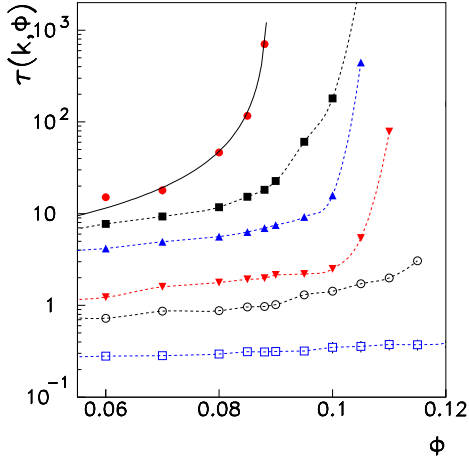}
\caption{(Color online)
Structural relaxation time $\tau_(k,\phi)$ as a function of
the volume fraction, for wave vector $k\sim 0.35$, $0.6$, $1.0$,
$2.0$, $3.0$, $7.0$ (from top to bottom). The full line is the
fitting curve: $\tau(k_{min},\phi) \sim (\phi_c-\phi)^{-f}$, with $f
\sim 1.22$. Dashed lines are eye
guides.}
\label{tau_phi}
\end{center}
\end{figure}

\begin{figure}
\begin{center}
\includegraphics[width=7cm]{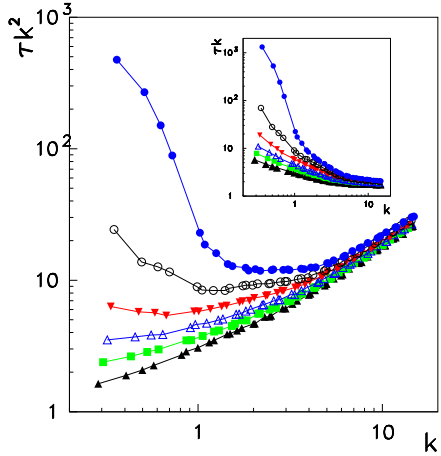}
\caption{(Color online)
{\bf Main frame}: $k^2 \tau(k,\phi)$ as a function of $k$,
for $\phi=~0.05$, $0.06$, $0.07$, $0.08$, $0.09$, $0.1$ (from bottom
to top). {\bf Inset}: $k \tau(k,\phi)$ as a function of $k$ for the
same volume fractions of main frame.
} \label{fig:tauk}
\end{center}
\end{figure}
From the self ISF we calculate the structural relaxation time,
$\tau(k,\phi)$, defined as the time for which $F_s(k,\tau(k))\simeq
0.1$. In Fig.\ref{tau_phi}, $\tau(k,\phi)$ is plotted for different
values of $k$ as a function of the volume fraction $\phi$. For
$k=k_{min}$, we find that  $\tau(k_{min},\phi)$ is well fitted by a
power law diverging at the gelation threshold with an exponent
$f\sim 1.22$.
Increasing $k$, no divergence is observed
at the threshold, signalling that no structural arrest occurs over
length scales less than the box size, $L$.

In Fig.\ref{fig:tauk} and in its inset we plot respectively
$k^2\tau(k,\phi)$ and $k\tau(k,\phi)$ as a function of the wave
vector for different volume fractions. The inset of
Fig.\ref{fig:tauk} shows that $\tau\sim 1/k$ for large wave vectors,
reflecting the ballistic diffusion for short times (see
Sect.\ref{diffusion}). Interestingly in the limit of small wave
vectors $k^2\tau$ does not tend to a constant. This unusual result
is essentially due to the fact that the the non-Gaussian parameter
\cite{Rahman}, $\alpha_2(t)=\frac{3 \Delta r^4(t)} {5(\Delta r^2(t))
^2}-1$, is non zero in the long time limit,
as discussed in details in Sect.\ref{diffusion}. In this case the
Gaussian approximation of the probability distribution of particle
displacements is not valid, and the self ISF $F_s(k,t)$ cannot be
written as a Gaussian even in the limit of small wave vector.

\begin{figure}
\begin{center}
\includegraphics[width=7cm]{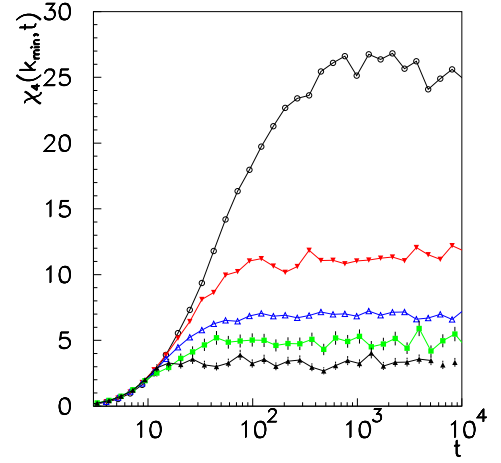}
\caption{(Color online)
Dynamical susceptibility, $\chi_4(k,t)$, as a function of
time for $k=k_{min}$ and different volume fractions
$\phi=0.05,~0.06,~0.07,~0.08,~0.09$ (from bottom to top).}
\label{fig5}
\end{center}
\end{figure}

We now analyze and discuss the behaviour of the fluctuations of the
self ISF, i.e. the dynamical susceptibility:
\begin{equation}
\chi_4(k,t)=N\left[\rule{0pt}{10pt}\lan |\Phi_s(k,t)|^2\ran-\lan
\Phi_s(k,t)\ran^2\right].
\end{equation}
In Fig.\ref{fig5} $\chi_4(k,t)$ is plotted for $k=k_{min}$ and
different volume fractions. Differently from the behavior typically
observed in glassy systems, we find that, for $\phi < \phi_c$,
$\chi_4(k,t)$ is a monotonically increasing function of the time
tending to a plateau in a time of the order of the relaxation time
$\tau(k_{min})$. The value of the plateau diverges as the mean
cluster size as the percolation threshold is approached
\cite{tiziana_prl}. For $\phi\ge \phi_c$ the system is out of
equilibrium, $\chi_4(k,t)$ continues increasing as a function of
time, without reaching any
asymptotic value within the simulation time.
We briefly discuss the main arguments explaining the above result,
presented in Ref \cite{tiziana_prl}, where it was in fact shown that,
for $k \to 0$ and $t\to\infty$, the dynamical susceptibility
$\chi_4(k,t)$ tends to the mean cluster size. We define
$\chi_{as}(k,\phi)\equiv \lim_{N\to\infty} \lim_{t\to\infty}
\chi_4(k,t)$. Being $\lim_{t\to\infty}\lan\Phi_s(k,t)\ran=0$, we
have
\begin{equation}
\chi_{as}(k,\phi)=
\lim_{N\to\infty}\frac{1}{N}\left[\sum_{i,j=1}^N\,C_{ij}(k)\right]
\label{chi4}
\end{equation}
where $C_{ij}(k)=\lim_{t\to\infty}\lan
e^{i\k\cdot(\r_i(t)-\r_j(t))}e^{-i\k\cdot(\r_i(0)-\r_j(0))}\ran
=|\lan e^{i\k\cdot(\r_i-\r_j)}\ran|^2$. Here we have used the fact
that, for large enough time $t$, the term
$e^{-i\k\cdot(\r_i(t)-\r_j(t))}$ is statistically independent from
$e^{-i\k\cdot(\r_i(0)-\r_j(0))}$, so that we can factorize the
thermal average. We separate the sum over connected pairs
($\gamma_{ij}=1$, i.e. pairs belonging to the same cluster) and
disconnected pairs ($\gamma_{ij}=0$, i.e. pairs belonging to
different clusters), so that: \begin{eqnarray}\label{eq:pairs}&&\chi_{as}(k,\phi)=\\
\nonumber &&\lim_{N\to\infty} \frac{1}{N}\left[\sum_{i,j=1}^N\,
\gamma_{ij}C_{ij}(k)\right]+
\frac{1}{N}\left[\sum_{i,j=1}^N\,(1-\gamma_{ij})C_{ij}(k)\right]
\end{eqnarray}
If particles $i$ and $j$ are not connected, for any fixed value of
$k>0$, the quantity $C_{ij}(k)$ is $O(1/N^2)$ \cite{nota2}. As there
are at most $N^2$ disconnected pairs, the second term of the r.h.s.
of Eq.(\ref{eq:pairs}) is  $O(1/N)$, and can be neglected in the
thermodynamical limit. \\For $\phi<\phi_c$, clusters will have at
most a linear size of order $\xi$, so that the relative distance
$|\r_i-\r_j|$ of connected particles will be smaller than $\xi$.
Therefore we have $\lim_{k\to 0}\gamma_{ij}C_{ij}(k)=\gamma_{ij}$
and
\begin{equation}
\lim_{k\to
0}\chi_{as}(k,\phi)=\lim_{N\to\infty}\frac{1}{N}\left[\sum_{i,j=1}^N\,\gamma_{ij}\right]
=S, \label{eq:approx}
\end{equation}
where $S$ is the mean cluster size. As shown in
Ref.\cite{tiziana_prl}, numerical data confirm this result.

\begin{figure}
\begin{center}
\includegraphics[width=7cm]{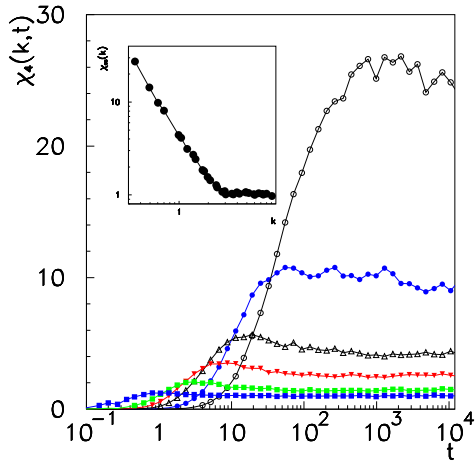}
\caption{(Color online)
{\bf Main Frame}: Dynamical susceptibility, $\chi_4(k,t)$,
as a function of time for $\phi=0.09$ and $k=0.35$, $0.61$, $0.99$,
$1.40$, $2.10$, $3.96$ (from top to bottom). {\bf Inset}: Asymptotic
values of the susceptibility, $\chi_{as}(k,\phi_c)$ as a function of
the wave vector $k$. Data are fitted with a power law $\sim
k^{-2.03\pm0.02}$, in agreement with the exponent $2-\eta$ of random
percolation.} \label{fig5_bis}
\end{center}
\end{figure}

In Fig.\ref{fig5_bis} $\chi_4(k,\phi)$ is plotted for $\phi=0.09$
and different wave vectors. For each value of the wave vector,
$\chi_4(k,\phi)$ reaches a plateau after a characteristic time of
the order of the relaxation time $\tau(k)$.
The asymptotic value $\chi_{as}(k,\phi_c)$ at low wave vectors
follows a scaling behavior as a function of $k$ (Inset of
Fig.\ref{fig5_bis}): at the transition threshold the exponent is
$2.03\pm0.02$, consistent, within the numerical accuracy, with the
prediction $2-\eta$ of random percolation \cite{stauffer}. This
result
shows that if one varies the wave vector $k$ (and $2\pi/k >
\sigma$) the dynamical susceptibility is able to detect the
self-similarity of the structure of the system due to the
percolation transition. Using scaling arguments \cite{tiziana_prl},
we can write $\chi_{as}(k,\phi) = k^{\eta-2}f(k\xi)$ where $f(z)$ is
a function, which tends to a constant for small $z$, whereas it
behaves as $z^{\gamma/\nu}$ for large values of $z$.
As shown in Ref.\cite{tiziana_prl}, data support this scenario.
All these results coherently show how in the present system the
asymptotic value of the dynamical susceptibility can be related to
the cluster size. Not only our results indicate that  the
percolation exponents can be measured in a direct way, by developing
techniques to measure the dynamical susceptibility, but they also
state that the asymptotic value of the dynamical susceptibility
plays the same role as the static scattering function near the
liquid gas critical point.

\section{Self Overlap and its fluctuations}
\label{sec:overlap} In the context of glassy systems, a
time-dependent order parameter was introduced \cite{59,60,61}, which
measures the number of ``overlapping'' particles in two
configurations separated by a time interval $t$,
\begin{eqnarray}
q(t)&&=\frac{1}{N}\int d\vec{r}_1
d\vec{r}_2\rho(\vec{r}_1,0)\rho(\vec{r}_2,t)
w(|\vec{r}_1-\vec{r}_2|) \nonumber\\
&&=\frac{1}{N}\sum_i\sum_j w(|\vec{r}_i(0)-\vec{r}_j(t)|),
\label{overlap}
\end{eqnarray}
where $\rho(\vec{r},t)= \sum_i \delta(\vec{r}-\vec{r}_i(t))$ is the
density in $\vec{r}$ at time $t$, and $w(|\vec{r}_1-\vec{r}_2|)$ is
an ``overlap'' function that is $1$ for $|\vec{r}_1-\vec{r}_2|\le a$
and zero otherwise \cite{nota_a}.

In Ref.\cite{glotzer} the authors separate $q$ into self and
distinct components, $q(t)=q_S(t)+q_D(t)$. The self part is given
by: \begin{equation} q_S(t)= \frac{1}{N}\sum_i w(|\r_i(0)-\r_i(t)|)
\label{eq:qs}\end{equation} which corresponds to terms of
Eq.(\ref{overlap}) with $i=j$, and measures the number of particles
that move less than a distance $a$ in a time interval $t$. In
Ref.\cite{glotzer} it was shown that on average the dominant term is
given by the self part.

\begin{figure}
\begin{center}
\includegraphics[width=7cm]{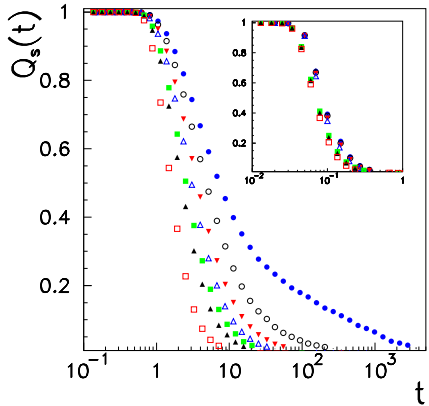}
\caption{(Color online)
{\bf Main frame}: Self overlap, $Q_S(t)$, for $a=~3$ and
different volume fractions from $\phi=0.02$, $0.05$, $0.06$, $0.07$,
$0.08$, $0.09$, $0.1$ (from bottom to top). {\bf Inset}: Self
overlap for the same values of $\phi$ of main frame and $a=0.15$ .}
\label{glotzer1}
\end{center}
\end{figure}

Here we measure $Q_S(t)=[\langle q_S(t)\rangle]$ for two choices of
$a$, $0.15$ and $3$, respectively corresponding to
$1/a >> k_{min}$ and $1/a \simeq k_{min}$.
$Q_S(t)$
is plotted in Fig.\ref{glotzer1} for different values of the volume
fraction. We have verified by numerical calculations that for small
enough $a$, the relevant contribution to $Q(t)=[\langle
q(t)\rangle]$ is given by $Q_S(t)$, since the probability that a
particle replaces within a radius $a$ another particle is small. For
all the values of $a$ and
of $\phi$ considered, $Q_S(t)$ at long times is well fitted
by a power law.

Another interesting method to investigate the spatially
heterogeneous dynamics, generally used in glassy systems, is the
measure of the dynamical susceptibility obtained by the fluctuations
of the time dependent overlap \cite{59,60,61}
$\overline{\chi}^Q_4(a,t)=N[\langle q(t)^2\rangle-\langle q(t)
\rangle^2],$ where $q(t)$ is given by Eq.(\ref{overlap}). In glassy
systems this quantity essentially presents the same features as the
fluctuations of the self ISF.

\begin{figure}
\begin{center}
\includegraphics[width=7cm]{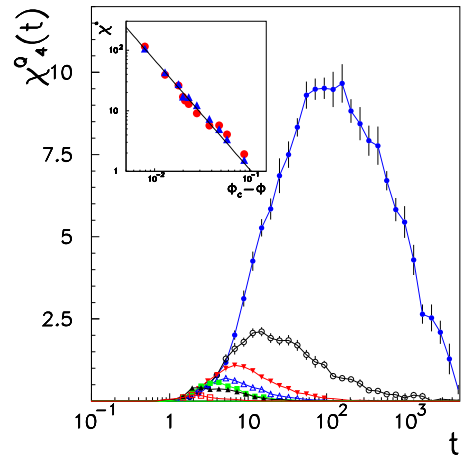}
\caption{(Color online)
{\bf Main frame}: Fluctuations of the self-overlap,
$\chi^Q_4(a,t)$, for $a=~3$ and
$\phi=0.02,0.05,0.06,0.07,0.08,0.09,0.1$ (from bottom to top). {\bf
Inset}: $10 \cdot \chi^Q_4(t^*)$ (circles) for $a=3$ and
$\chi_{as}(k_{min},\phi)$ (triangles) as a function of
$(\phi_c-\phi)$.} \label{glotzer2}
\end{center}
\end{figure}

Here we measure the fluctuations of the self part of the overlap:
\begin{equation}
\chi^Q_4(a,t)=N [\langle q_S(t)^2\rangle-\langle q_S(t)
\rangle^2],\label{eq:chiqs}
\end{equation}
for different choices of $a$, ranging from $0.15$ to $3$. In
Fig.\ref{glotzer2} we plot $\chi^Q_4(a,t)$ for $a=3$ and different
values of $\phi$.

\begin{figure}
\begin{center}
\includegraphics[width=7cm]{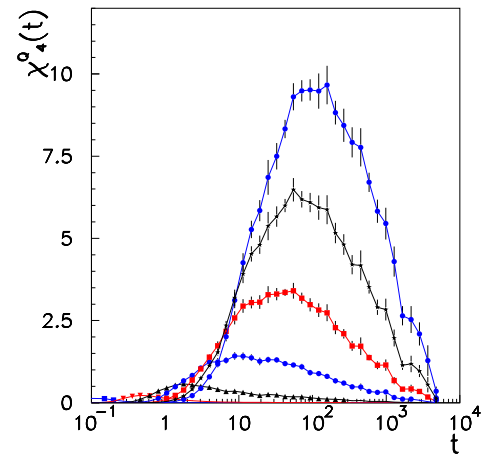}
\caption{(Color online)
{\bf Main frame}: Fluctuations of the self-overlap,
$\chi^Q_4(a,t)$, for $\phi=0.1$ and $a=0.15,0.5,1,1.5,2,2.5,3$ (from
bottom to top).
} \label{picco}
\end{center}
\end{figure}

We see that differently from the fluctuations of the self ISF, here
$\chi^Q_4(a,t)$ displays a peak, whose value increases and diverges
as the the gel transition is approached. Indeed, the value of the
peak differ from the value of the plateau $\chi_{as}(k_{min})$ only
for a constant factor (see inset of Fig.\ref{glotzer2}) and
therefore scales with the same exponent of the mean cluster size
$\gamma$ \cite{note}. Even if the long time limit of $\chi^Q_4(a,t)$
is strongly different from the one observed in $\chi_4(k,t)$, both
fluctuations manifest a strong dependence on length scale. In fact,
the peak of $\chi^Q_4(a,t)$ strongly decreases as $a$ decreases (see
Fig.\ref{picco}). This may be interpreted as a sign that clusters of
bonded particle dominate the dynamics.

Our data and these considerations suggest that heterogeneities
detected by $\chi^Q_4(a,t)$ are due to the presence of clusters.
However, despite the permanent nature of clusters, fluctuations of
the self overlap $\chi^Q_4(a,t)$ decay to zero in the long time
limit. This is due to the form of the overlap function
$w(|\r_i(0)-\r_i(t)|)$, which is zero when a particle has moved a
distance greater than $a$. Therefore two particles in the same
cluster will contribute to $\chi^Q_4(a,t)$, if the center of mass of
the cluster has moved a distance less than $a$. In fact, when the
cluster moves a distance larger than $a$, due to the form of the
overlap function, the contribution to $\chi^Q_4(a,t)$ vanishes.
Therefore we expect that for $1/a \simeq k_{min}$ the peak is
proportional to the mean cluster size, and occur at a time $t^*$ of
the order of the time in which the center of the typical cluster of
dimension $\xi$ has moved a distance of the order of $a$.

\section{Mean square displacement and the non-Gaussian parameter}
\label{diffusion}

\begin{figure}
\begin{center}
\includegraphics[width=7cm]{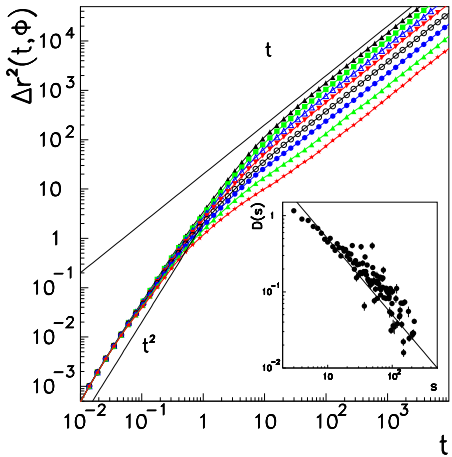}
\caption{(Color online)
{\bf Main frame}: Mean square displacement for $\phi=~0.05$, $0.06$,
$0.07$, $0.08$, $0.09$, $0.1$, $0.11$, $0.12$. {\bf Inset}:
Diffusion coefficient, $D(s)$, as a function of the cluster size $s$
for $\phi=\phi_c$.} \label{rsq}
\end{center}
\end{figure}

Finally we have measured the mean square displacement (MSD)
\begin{equation}
\Delta r^2(t)=\frac{1}{N}\sum_{i=1}^N\left[\langle |\vec{r}_i(t)-
\vec{r}_i(0)|^2\rangle\right],
\end{equation}
where $\vec{r}_i(t)$ is the position of the $i$-th particle at the
time $t$.
In the main frame of Fig.\ref{rsq} the MSD is shown for different
volume fractions. Due to the Newtonian dynamics, we find at very
short time a ballistic behaviour, $\Delta r^2(t)\propto t^2$,
followed by a crossover to a diffusive regime  $\Delta r^2(t)\propto
t$. The long time diffusive regime is always recovered for all the
volume fractions considered, indicating that even at the percolation
threshold this quantity is dominated by free motion of particles or
clusters. Accordingly, no divergence of the inverse diffusion
coefficient is found at the percolation threshold, where the
numerous small size clusters continue to diffuse into the large mesh
of the spanning cluster.

We have also evaluated the MSD of the clusters and extracted
their diffusion coefficient
as a function
of the cluster size $s$. In particular we obtained that for
$\phi=\phi_c$, $D(s)$ for large $s$ is fitted by a power law
$s^{-h}$ with $h=1.0 \pm 0.1$ (see inset of Fig.\ref{rsq}).
Following \cite{ema_EPJ} we expect $D(s)\sim 1/s^{(d-2+f/\nu)/d_f}$,
where $d=3$ is the Euclidean dimension, $f$ is the exponent which
gives the divergence of the viscosity, $\nu\sim 0.88$ is the
critical exponent which gives the divergence of the connectedness
length (see Appendix), and $d_f\sim 2.4$ is the fractal dimension of
the spanning cluster at the threshold (see Appendix). Using these
values we obtain a prediction for the exponent, which gives the
divergence of the viscosity at the threshold $f=\nu(hd_f-d+2)\sim
1.23$ in agreement within the errors with our data for the
structural relaxation time (see Sect.\ref{sec:self} and
Fig.\ref{tau_phi}).

\begin{figure}
\begin{center}
\includegraphics[width=7cm]{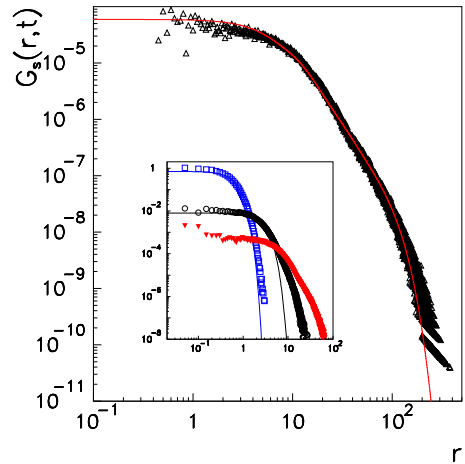}
\caption{(Color online)
{\bf Main frame}: The self part of the Van-Hove
distribution for $\phi=0.09$ and time $t=~1285.02$. The continuous
line is obtained from the diffusion coefficient of clusters using
Eq.(\ref{self-van-hove}). {\bf Inset}: The  self part of the
Van-Hove distribution for $\phi=0.09$ and different times $t=0.469$,
$6.739$, $93.199$ (from left to right). The lines are Gaussian
fitting functions.} \label{fig:gr}
\end{center}
\end{figure}

In order to characterize the displacement of particles we have
calculated the self part of the Van-Hove function \cite{hansen}: \be
G_s(r,t)=\frac{1}{N}\left[\langle\sum_{i=1}^N\delta
(r-|\r_i(t)-\r_i(0)|) \rangle\right].\ee If the motion of particles
is diffusive
with a diffusion coefficient D,
$G_s(r,t)=(1/4\pi Dt)^{3/2}e^{(-r^2/4Dt)},$ where $r$ is the
distance traveled by a particle in a time $t$. In the inset of
Fig.\ref{fig:gr} we plot the self van Hove function for a fixed
volume fraction at different times. Our results indicate that for
short times and short distances the function is well fitted by a
Gaussian. For long distances and long times, the van Hove function
is well fitted by an exponential decay. An exponential decay has
been observed in different glassy systems for intermediate times
\cite{stariolo}.

The deviation from the Gaussian distribution at long times, observed
in our system, indicates that some particles move faster than
others, due to the presence of clusters. Particles belonging to
different clusters have a different diffusion coefficient depending
on the cluster size. As a consequence we suggest that, in the
diffusive regime, $G_s(r,t)$ does not have a Gaussian form, but it
is instead given by a superposition of Gaussians
\begin{equation}
G_s(r,t)=\sum_s s n_s \left(\frac{1}{4\pi
D(s)t}\right)^{3/2}e^{-r^2/4D(s)t}, \label{self-van-hove}
\end{equation}
where $D(s)$ is the diffusion coefficient of cluster of size $s$ and
$n_s$ is the cluster size distribution.

In Fig.\ref{fig:gr} we compare our data with $G_s(r,t)$ calculated
using Eq.(\ref{self-van-hove}) and $D(s)$ obtained from the
simulations. As we can see in figure, data are in good agreement
with our predictions, provided that time is sufficiently long for
clusters diffusing with diffusion coefficient $D(s)$.

In agreement with this finding, the non-Gaussian parameter, which
is a measure of the departure from the Gaussian behaviour of the
probability distribution of the particle displacements, does not go
to zero at long times. The non-Gaussian parameter is defined as
\cite{Rahman}:
\begin{equation}
\label{eq:alfa} \alpha_2(t)=\frac{3 \Delta r^4(t)} {5 (\Delta
r^2(t)) ^2}-1,
\end{equation}
where $\Delta r^4(t)=\frac{1}{N}\sum_{i=1}^N\left[\langle
|\vec{r}_i(t)- \vec{r}_i(0)|^4\rangle\right]$ and it
is zero if the probability distribution of the particle
displacements
is Gaussian.

In glassy systems \cite{kob_PRL_97}, (i) on the time scale
at which the motion of the particles is ballistic, $\alpha_2$ is
zero; ii) upon entering the time scale of the $\beta$ relaxation,
$\alpha_2$ starts to increase; iii) on the time scale of the
$\alpha$ relaxation, $\alpha_2$ decreases to its long time limit,
zero. The maximum value of $\alpha_2$ increases with decreasing
temperature, signalling that the dynamics becomes more heterogeneous.

In the present model for permanent gels, we find that i) as in
glasses, $\alpha_2$ is always zero on the time scale at which the
motion of the particles is ballistic; ii) it tends in the long time
limit to a plateau value, which increases with increasing volume
fraction; iii) at low volume fraction, $\alpha_2$ has a maximum at
intermediate times, which disappears upon approaching the gelation
threshold; iv) no critical behaviour is observed at the gelation
threshold.

Within our interpretation the asymptotic value of the non-Gaussian
parameter, using Eq.(\ref{self-van-hove}) may be written in the following form:
\begin{equation}
\alpha^{as}_2=\frac{\sum_s s n_s D^2(s)}{(\sum_s s n_s D(s))^2}-1=
\frac{\overline{ D ^2} -\overline{D} ^2}{\overline{D} ^2},
\end{equation}
where, for each bond configuration, $\overline{(\dots)}$ is the
average over the cluster distribution. We have verified that
$\alpha^{as}_2$ coincides with $\frac{\overline{ D ^2} -\overline{D}
^2}{\overline{D} ^2}$ within the errors. Hence our results indicate
that the non-Gaussian parameter tends to a plateau given by the
ratio of two quantities, which both have no critical behaviour at
the percolation threshold. In summary, as the fluctuations of the
self ISF, the non-Gaussian parameter does not decay to zero in the
long time limit, due to the presence of permanent clusters. However,
the main contribution to $\alpha_2$ comes from the numerous finite
clusters (the bigger the cluster, the lower its diffusion
coefficient $D(s)$, and consequentially its contribution to
$\alpha_2$), so that no criticality approaching the percolation
threshold is observed in the non Gaussian parameter.

\begin{figure}
\begin{center}
\includegraphics[width=7cm]{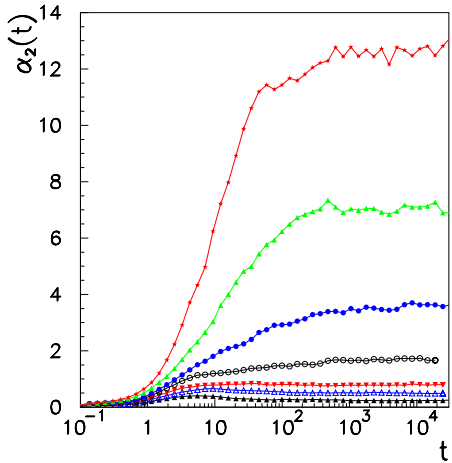}
\caption{(Color online)
Non-Gaussian parameter, $\alpha_2(t)$ as a function of time
$t$ for $\phi=0.05$, $0.06$, $0.07$, $0.08$, $0.09$, $0.1$, $0.12$.}
\label{alfa}
\end{center}
\end{figure}

\section{Conclusion}
\label{conclusion}

We have presented a molecular dynamics study of a model for
permanent gels and investigated its static and dynamical properties.
Usually the sol-gel transition, marked by the divergence of
viscosity and the onset of an elastic modulus, is interpreted in
terms of the appereance of a percolating cluster of monomers linked
by bonds \cite{flo, deg, stauffer}. While the viscosity and the
elastic modulus can be measured directly, usually the experimental
determination of percolative properties needs the manipulation of
the sample (for a review see \cite{adconst} and references therein),
i.e. the sample is dissolved in a known quantity of solvent in such
a way that each cluster is separated from the others. For the first
time our results identify the thermodynamical observable associated
with the cluster properties in a gelling system, and, via the
measure of the fluctuations of the self ISF, allow to obtain the
critical exponents without such a manipulation of the sample.

In our model the formation of permanent bonds between the particles
leads to a percolation transition in the universality class of
random percolation. The percolation threshold coincides with the
gelation threshold, marked by the slowing down of dynamics on length
scale of the whole system. We have found that the behavior of the
self ISF in the sol phase and near the threshold is in agreement
with typical experiments on gelling systems. In chemical gels the
onset of a stretched exponential decay is typically associated to
the wide cluster size distribution close to the gelation threshold,
producing a wide distribution of relaxation times. At the
percolation threshold, the longest relaxation time diverges due to
the critical growing of the percolation correlation length,
producing a long time power law decay. Our results  confirm this
picture but new insights are obtained with a study of the dynamical
heterogeneities, in terms of fluctuations of different correlation
functions. In the present model for permanent gels, the fluctuations
of the self-overlap present always a peak, whereas the fluctuations
of the self ISF are monotonically increasing with time. Differently
from glassy systems, the fluctuations of the self ISF tend in the
long time limit to a plateau, whose value, for the lowest wave
vector, coincides with the mean cluster size. The behavior of the
non-Gaussian parameter as a function of time is qualitatively
similar: in the long time limit it reaches a plateau, due to the
contribution of particles belonging to different clusters with a
size dependent diffusion coefficient. Nevertheless, the value of the
plateau does not diverge at the gelation transition, being dominated
by the presence of small clusters with finite diffusivity.
\begin{figure}
\begin{center}
\includegraphics[width=7cm]{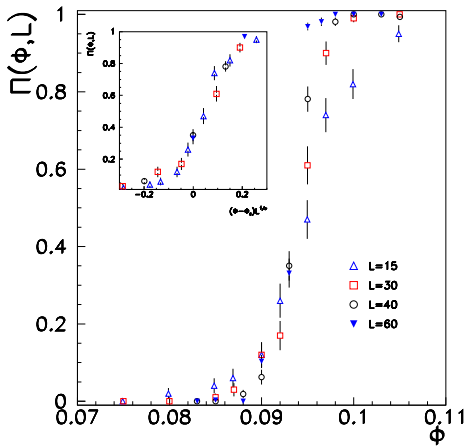}
\caption{(Color online)
{\bf Main frame}: Percolation probability $\Pi(\phi,L)$ as
a function of the volume fraction $\phi$ for boxes of size $L=15$,
$30$, $40$, $60$. {\bf Inset}: Data collapse obtained plotting
$\Pi(\phi,L)$ versus $(\phi-\phi_c)L^{1/\nu}$ with $\nu=0.88$ and
$\phi_c=0.09$.} \label{fig1}
\end{center}
\end{figure}
\begin{figure}
\begin{center}
\includegraphics[width=7cm]{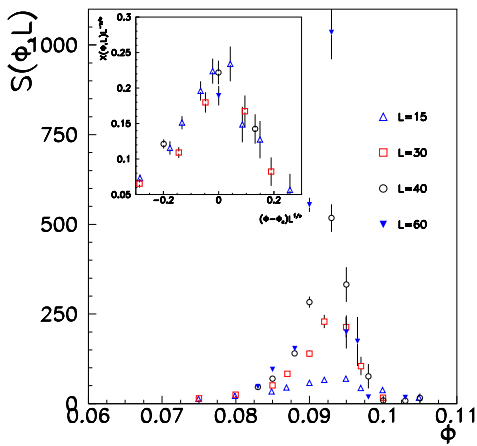}
\caption{(Color online)
{\bf Main frame}: Mean cluster size $S(\phi,L)$ as a
function of the volume fraction $\phi$ for boxes of the same sizes
of Fig. \ref{fig1}. {\bf Inset}: Data collapse obtained plotting
$S(\phi,L)L^{-\gamma/\nu}$ versus $(\phi-\phi_c)L^{1/\nu}$ with
$\nu=0.88$, $\phi_c=0.09$ and $\gamma=1.85$.} \label{fig2}
\end{center}
\end{figure}

This study has shed some light on the differences between the
dynamics and the dynamical heterogeneities in glasses and chemical
gels. We have been able to clarify that, when clusters of bonded
particles are present, different time correlators can deliver very
different information whereas in the studies on glasses they are
often used interchangeably. On this basis, these findings have
interesting implications for the study of gels due to non-permanent
bonds, as in the case of colloidal gels. In fact, our study also
indicate a possible way to discriminate between a gel-like behaviour
and a glass-like behaviour in these systems. Our results strongly
suggest that, if heterogeneities are due to clusters of particles
connected by permanent (or persistent) bonds, as in permanent (or
colloidal) gels, the behavior of the ``time-dependent order
parameter'', whose fluctuations reveal the presence of
heterogeneities in the dynamics, may be quite different. However
both quantities show a strong length scale dependence (both strongly
decrease as $k$ or $1/a$ increase), which seems to be the distinct
sign of (permanent or colloidal) gelation compared with the
(attractive or repulsive) glass transition \cite{decandia_prep}.
This result is confirmed by a recent work \cite{jstat}, where it has
been found that, in a model for colloidal gels, at low volume
fraction, the fluctuations of the self ISF for small wave vector
display a dependence on time, which is dramatically different from
the one found at higher volume fraction \cite{charb,cipelletti} in
the glassy regime. As a final remark, it is interesting to note that
in the model here discussed the dynamical susceptibility is similar
to that observed in a spin glass model with quenched interactions
\cite{lattice-gas}, suggesting a possible common description of the
phase transition involved, as also proposed elsewhere
\cite{gel_dyn_theo}.

{\it Aknowledgements:} The research is supported by the EU Network
Number MRTN-CT-2003-504712, INFM-PCI and S.Co.P.E.

\appendix
\section{Percolation transition}
\label{gelation} \begin{figure}
\begin{center}
\includegraphics[width=7cm]{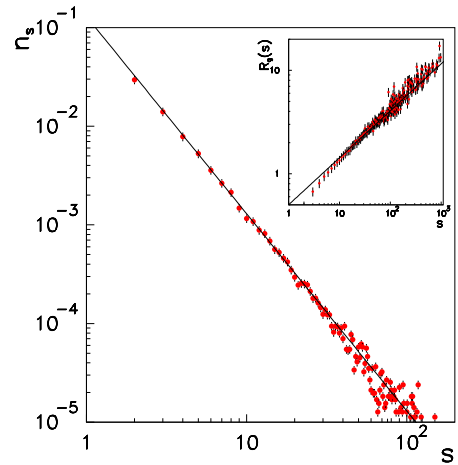}
\caption{(Color online)
{\bf Main frame}: Average number of clusters (per particle)
with mass$s$ versus $s$at $\phi_c$ for $L=40$. The full line is a
power law, $s^{-\tau}$, withfitting parameter $\tau=2.1$. {\bf
Inset}: Radius of gyration, $R_g$, as a function of the mass $s$ of
clusters at $\phi_c$ for $L=40$. The full line is a power law,
$s^{1/d_f}$, with fitting parameter $d_f=2.4$. } \label{fig3}
\end{center}
\end{figure}
In this Appendix, with a finite size scaling
analysis, the percolation threshold and the critical exponents are
obtained. We find that the percolation of permanent bonds,
corresponding to the sol-gel transition \cite{flo,deg}, is in the
universality class of random percolation.

Varying the volume fraction $\phi$, we have measured the percolation
probability, $\Pi(\phi)$ (defined as the average number of
configurations where a percolating cluster is found), the cluster
size distribution, $n_s$, and the mean cluster size $S(\phi)= \sum
s^2 n_s / \sum s n_s$. For each volume fraction we have used
simulation boxes of different size $L$ and, from a standard finite
size scaling analysis \cite{stauffer}, we have obtained the
percolation threshold $\phi_c$, and the critical exponents $\nu$
(which governs the power law divergence of the connectedness length
$\xi\sim|\phi-\phi_c|^{-\nu}$ as the transition threshold is
approached from below) and $\gamma$ (which governs  the power law
divergence of the mean cluster size $S\sim
|\phi-\phi_c|^{-\gamma}$). The percolation threshold and the
critical exponents obtained from the data showed in Fig.s \ref{fig1}
and \ref{fig2}, are respectively $\phi_c=0.09\pm0.01$,
$\nu=0.88\pm0.05$ and $\gamma=1.85\pm0.05$. The cluster size
distribution $n_s$ for $\phi=\phi_c$, shown in main frame of Fig.
\ref{fig3}, follows a power law behavior $n_s\sim s^{-\tau}$ with a
Fisher exponent $\tau=2.1\pm0.2$.

Finally in the inset of Fig.\ref{fig3}
the radius of gyration $R_g$ as a function of the mass $s$ of clusters
is plotted. The data are well fitted by a power law with
exponent $1/d_f=0.42\pm0.03$ (full line in figure)
which gives $d_f=2.4\pm0.1$ in agreement with the
fractal dimension of the random percolation clusters in $3d$, $d_f\simeq2.5$.
The measured values of the critical exponents
satisfy the hyper-scaling relations ($2\beta+\gamma=\nu d$,
$d_f=d-\beta/\nu$, and $\tau=2+(d-d_f)/d_f$ \cite{stauffer}),
and are in good agreement with those
of the $3d$ random percolation  ($\nu=0.88$, $\gamma=1.80$ and
$\tau=2.18$ \cite{stauffer}).


\begin{thebibliography}{99}
\bibitem{nota} Increasing $k$ would reduce the maximum extension further
but would require a reduction in $\Delta t$.
\bibitem{nota2}
We can write
\begin{equation}
\lan e^{-i\k\cdot(\r_i-\r_j)}\ran
=\frac{1}{N}\int\!d^3\r\,e^{-i\k\cdot\r}\,\rho [h_{ij}(\r)+1],
\label{meikr}
\end{equation}
where $\rho= N/V$, $h_{ij}(\r) + 1 = g_{ij}(\r)$ and
$(1/V)g_{ij}(\r)$ gives the probability density of finding the
particle $i$ in $\r$, given the particle $j$ in the origin. For
disconnected pairs, in the thermodynamic limit ($N\to\infty$ and
$L\to\infty$ leaving the density $\rho$ constant) the integral in
Eq.(\ref{meikr}) remains finite for any finite fixed $k$, so that
the l.h.s. of Eq.\ (\ref{meikr}) is  $O(1/N)$. In fact the first
term, the Fourier transform of the correlation function
$h_{ij}(\r)$, is finite when $L\to\infty$ since $h_{ij}(\r)$ decays
to zero at a finite distance. The second term, the modulus of the
Fourier transform of $1$, is not larger than $8/|k_xk_yk_z|$, and
consequently remains finite when $L\to\infty$. The quantity
$C_{ij}(k)$, being the square modulus of the l.h.s. of Eq.\
(\ref{meikr}), is therefore $O(1/N^2)$.
\bibitem{nota_a} In the simulations this is implemented by choosing
$\omega(r)=a^{12}/(a^{12}+r^{12})$. In glassy systems the parameter
$a$ is usually chosen $\sim 0.3$. The reasons of this choice are
discussed in Ref.\cite{glotzer}.
\bibitem{note}
Neglecting the correlation between particles belonging to different
clusters,
and supposing that the displacement of each cluster at the time $t$
depends only on its size $s$, $\chi^Q_4(a,t)$ can be written for
long enough times in the following way:
\begin{equation}
\chi^Q_4(a,t)\sim \sum_s n_s s^2 \left(\langle w^2(r_{CM}(s,t))
\rangle- \langle w(r_{CM}(s,t))\rangle^2\right), \nonumber
\end{equation}
where $r_{CM}(s,t)$ is the displacement of any cluster of size $s$
at the time $t$. In principle $f(s,t)=\langle w^2(r_{CM}(s,t))
\rangle- \langle w(r_{CM}(s,t))\rangle^2$ depends on the cluster
size $s$. If $f(t,s)$, for a given $a$, displays, at a time $t^*$,
a peak whose value does not depend on the cluster size $s$, we
obtain that $\chi^Q_4(t^*)$ is proportional to the mean cluster
size.
\end{thebibliography}
\end{document}